# An overview and perspective on social network monitoring

WILLIAM H. WOODALL[1], MENG J. ZHAO[1], KAMRAN PAYNABAR[2], ROSS SPARKS[3] and JAMES D. WILSON[4]

1 *Department of Statistics, Virginia Tech, Blacksburg, VA 24061-0439*
bwoodall@vt.edu and jmz167@vt.edu

2 *H. Milton Stewart School of Industrial and Systems Engineering, Georgia Institute of Technology, Atlanta, GA 30332* kamran.paynabar@isye.gatech.edu

3 *CSIRO Computational Informatics, Locked Bag 17, North Ryde NSW 1670 Australia* Ross.Sparks@csiro.au

4 *Department of Mathematics and Statistics, University of San Francisco, San Francisco, CA 94117-1080* jdwilson4@usfca.edu

In this expository paper we give an overview of some statistical methods for the monitoring of social networks. We discuss the advantages and limitations of various methods as well as some relevant issues. One of our primary contributions is to give the relationships between network monitoring methods and monitoring methods in engineering statistics and public health surveillance. We encourage researchers in the industrial process monitoring area to work on developing and comparing the performance of social network monitoring methods. We also discuss some of the issues in social network monitoring and give a number of research ideas.

## 1. Introduction

There has been an increasing amount of research on the monitoring of social networks. An overview of methods was given in a recent review paper by Savage et al. (2014) who listed applications including the detection of important and influential network participants, the detection of clandestine organizational structures, and the detection of fraudulent or predatory activity. One of our primary contributions is to add to the discussion of Savage et al. (2014) by including additional network monitoring papers and discussing the various methods in the context of the considerable amount of related work in industrial process monitoring and public health surveillance. Social network monitoring methods are often illustrated using terrorist networks like the al Qaeda network (see Figure 1) or social networks such as that based on Enron e-mail communications (see Figure 2).

The basic idea in social network monitoring is to detect sudden changes in the behavior of a subset of the individuals in the network. Significant increases in the communication levels of the entire network, of smaller sub-networks or of individuals are often of primary interest in applications, where global changes are typically the easiest to detect. In some cases, however, decreases in communication levels may be of interest. Savage et al. (2014) referred to regions of the network with structure differing from that expected under normal conditions as *anomalies*. Of course, to formalize what is meant by an anomaly, there must be an operational definition of the normal conditions. The definition of an anomaly would likely vary from application to application. Networks are expected to evolve over time, however, so it would be unusual to have interest in detecting that any change, however small, has occurred. The focus is usually on detecting sudden large changes in the structure of some portion of the network.

We assume that there are $n$ individuals in the network to be monitored. These individuals could refer to people, e-mail addresses, or other entities. We assume that we are collecting network data aggregated over some time period to give, for example, daily or weekly data, with $m$ time periods of data in a baseline sample. For each time period $t$, $t = 1, 2, \ldots$, we have information on the communication level between individual $i$ and individual $j$, say $c_t(i, j)$, $i, j = 1, \ldots, n$, where $i$ is not equal to $j$. Most often we are interested in the number of communications between individuals $i$ and $j$. Alternatively, $c_t(i, j)$ may be an indicator variable indicating whether or not there was at least one contact between $i$ and $j$, or whether some other criterion on the level of communication between these two individuals was met. In the social network change detection literature, the numbers of contacts between pairs of individuals are frequently modeled by some variant of the Poisson distribution; whereas, Bernoulli random variables are typically used to model indicator variables. Communication levels can be quantified as directed or undirected. With directed data, $c_t(i, j)$ reflects only communications between individuals $i$ and $j$ that were initiated by individual $i$; whereas, with undirected data, communications are considered mutual, namely $c_t(i, j) = c_t(j, i)$. There can be a substantial loss of information in transforming directed to undirected data, or in representing communication counts by binary indicator variables. Indeed, with undirected data it is not possible to study how contacts propagate through the network. Generally, as discussed by Schuh et al. (2013), greater levels of data aggregation result in greater losses of information and poorer process monitoring performance.

The values $c_t(i, j)$ can be placed into row $i$ and column $j$ of a matrix, say $\boldsymbol{C}_t$, $t = 1, 2, \ldots$ . The matrix $\boldsymbol{C}_t$ is typically referred to as the *adjacency matrix* or *graph* corresponding to the social network at time $t$. These matrices are usually quite sparse and assumed to have diagonal elements set to zero so that the graph contains no self-loops. Note that if the data are undirected, then the matrix $\boldsymbol{C}_t$ is symmetric. Thus the network monitoring problem can be framed as the detection of certain types of changes in matrices of indicator variables or counts over time. This is a broad generalization of the framework usually considered in the many papers on the monitoring of Bernoulli or count data. The vast majority of the methods for such data studied in the literature on statistical process monitoring are univariate and thus could be applied directly only to a network consisting of two individuals.

As reviewed by Szarka and Woodall (2011), there has been much research on the monitoring of sequences of Bernoulli data. Aside from its diagonal of zero elements, $\boldsymbol{C}_t$ will be a matrix of Bernoulli random variables in network monitoring applications in which $c_t(i, j) = 1$ if there was at least one contact between individuals $i$ and $j$ or some other criterion was met, and 0 otherwise. This represents a substantial multivariate generalization of the usual univariate framework.

The monitoring of a single stream of Poisson-distributed data has been widely studied. He et al. (2012) provided a review of methods for monitoring a zero-inflated Poisson distribution. Purdy et al. (2015) reviewed methods for monitoring non-homogeneous Poisson processes. Monitoring with multivariate Poisson vectors has been studied by Laungrungrong et al. (2011) and He et al. (2014), among others, but no one has studied the monitoring of matrices of Poisson counts in the industrial statistics literature.

Li et al. (2012) and Yashchin (2012) proposed methods for monitoring categorical data which can be considered in some cases to consist of matrices of counts, but not with the same matrix structure or the same objectives as in network monitoring. It is frequently assumed in the study of public health surveillance methods, however, that each sample of disease incidence counts consists of a set of assumed Poisson random variables that could be viewed as components of a matrix. Sometimes these counts apply to a rectangular set of sub-regions of a larger region of interest. Scan methods, such as those of Kulldorff (2001), are frequently used with this type of data to detect clusters of contiguous sub-regions where the Poisson rate seems significantly higher than expected. A primary difference between this problem and that of network monitoring is that location can provide a natural ordering of the sub-regions in public health applications. There is usually no natural ordering for individuals in a network.

There has been considerable work on the monitoring of computer networks. See, for example, Neil et al. (2013). In their review, Savage et al. (2014) pointed out that the structure of social network data is usually different from that of computer networks and that the objectives are typically different. In addition, computer network monitoring involves considerably more data collected at a much higher frequency and with much more pronounced periodic patterns than with social network data. Newman and Park (2003) provided a discussion of the differences between these two types of networks. The social network monitoring literature seems to have been developed somewhat independently of the computer network monitoring literature.

In the Appendix we give a brief introduction to social network terminology. Section 2 contains our review of social network monitoring methods. In Section 3 we discuss some issues related to network monitoring. Conclusions and a number of research topics are given in Section 4.

# 2. Monitoring methods

In this section we briefly describe some of the recent methods proposed for monitoring social networks and relate them to methods in the area of statistical process monitoring. We use categories corresponding roughly to those used by Savage et al. (2014). These four categories were also used in the review paper by Unkel et al. (2011) to classify prospective public health surveillance methods. We assume that the reader has some familiarity with statistical process monitoring methods. For more information on these methods, we recommend Montgomery (2013).

1. *Control chart and hypothesis testing methods*

We believe that concepts and methods in statistical process monitoring can be used to greater advantage in social network monitoring. One of these concepts, the distinction between the retrospective analysis of baseline data (Phase I) and methods for prospective on-going monitoring (Phase II), is discussed in Section 4.1.

In their papers, McCulloh and Carley (2008a, b), McCulloh et al. (2008) and McCulloh and Carley (2011) used monitoring methods such as the cumulative sum (CUSUM) and exponentially weighted moving average (EWMA) charts to detect changes in the network as a whole. For information on these two types of charts, we recommend Hawkins (2014). They focused on detecting changes in the communication behavior in military units. Global network metrics such as average closeness and average betweenness were used as time series input to the charting methods, but it was pointed out that node or sub-network metrics could have been used instead. McCulloh and Carley (2011) stated that five or more network graphs should be used to establish a baseline. Current research by Saleh et al. (2016) and others show, however, that many more network graphs would have to be available in order to estimate the baseline parameters so that the resulting control chart performance would be reliable.

Azarnoush et al. (2016) proposed a method to detect changes in the behavior within and between specified sub-networks with the incorporation of covariate information. For example, in a university environment the sub-networks could correspond to departments and faculty rank could serve as a covariate. The authors modeled the probabilities of contacts between pairs of individuals in the network using a logistic regression model with sub-network membership and covariate data on the individuals used as explanatory variables. A likelihood ratio test was proposed to detect changes in the logistic regression model fit with each new graph. The authors proposed three approaches. One is referred to as the *static reference approach* where each new graph is compared to those in a fixed baseline Phase I sample. In the *dynamic reference approach*, each incoming graph is compared to all previous graphs. If there is no signal of an anomaly then the current graph is entered into the baseline for the next graph to be observed. The third approach is referred to as the *dynamic reference sliding window approach* where the current graph is compared to only the most recent $q$ graphs, where $q$ is the size of the moving window. Azarnoush et al. (2016) stated that the choice of approach depends on the objective of monitoring, but we see the moving window approach as generally being the most useful because networks tend to evolve over time.

Some types of anomalies cannot be detected with the logistic regression method of Azarnoush et al. (2016). In some cases the number of contacts within a specified group within the network can be redistributed into any configuration without affecting the estimated regression coefficients or the likelihood ratio test. We note that checking for changes in a logistic regression model over time falls into the category of profile monitoring. Yeh and Huang (2011) reviewed some relevant methods for determining whether or not a logistic regression model has changed over time.

Azarnoush et al. (2016) used simulation to compare their methods to those of McCulloh and Carley (2011), where the latter methods are based on global network metrics without the incorporation of the available covariate information. It would have been a fairer comparison, however, to base the McCulloh and Carley (2011) methods on metrics corresponding to the activity within each of the two assumed categories of individuals.

Miller et al. (2013) proposed a method with assumptions quite similar to those of Azarnoush et al. (2016) to detect specified types of network changes. Miller et al. (2013) used a log-linear model for the probabilities of connections between pairs of individuals, however, because the reduced amount of computation allows their method to be used with much larger networks. The monitoring approach of Miller et al. (2011, 2013) is based on eigenvalues of modularity matrices, proposed by Newman (2006) for finding community structure in networks. The modularity matrix is the difference between $C_t$ and the expected value of $C_t$ assuming that edges occur independently. It can be thought of as a residuals matrix. Miller et al. (2011, 2013) considered a window of network snapshots to calculate the differences between observed and expected adjacency matrices. The differences between the matrices were weighted with filter coefficients based on the assumed known signal model. Instead of only choosing the first eigenvector of the resulting matrix, they picked the first two and projected the modularity matrix onto the corresponding space. They assumed if there is no change, projected values should be randomly scattered (not clustered) in any arbitrarily defined quadrant. In order to test this hypothesis, they use a contingency test statistic in a 2 by 2 table defined by the quadrant. If the test statistic is large, there is evidence of change.

Miller et al. (2011) assumed that there is a known signal model, where the anomalous subgraph behavior of interest is known, but its position within the background is not. Their matched filter approach is very similar to the cuscore approach of Box and Ramirez (1992), which Apley and Chin (2007) showed was not effective at detecting delayed process shifts.

In their first simulation Miller et al. (2011) assumed that an anomalous subgraph density is fixed, but the edges changed with each sample. In this case their approach gets better and better as the window size increases. This is appropriate for hypothesis testing, but large window sizes are not efficient for process monitoring because of the buildup of inertia. As discussed by Woodall and Mahmoud (2005), inertia can slow down the detection of delayed process changes. Miller et al. (2011) assumed in their second simulation that the density of the anomalous subgraph increased linearly over 32 samples. The filter coefficients were then set to be linearly decreasing with age from 1 to 0 over a window of size 32. The problem with their matched filter approach is that one does not know when the signal (anomaly) will occur. Miller et al. (2011, 2013) assumed a hypothesis testing framework, not on-line continuous Phase II monitoring, and evaluated their methods using ROC curves assuming that any anomalies occur immediately.

## 2. Bayesian methods

Heard et al. (2010) proposed a two-stage Bayesian approach to anomaly detection. Their goal was to detect anomalous communication levels between pairs of individuals. Once these pairs are identified, they are used to form a sub-network that can then be analyzed for anomalous behavior. They assumed either a Poisson conditional distribution or a hurdle Poisson conditional distribution for the counts of contacts between pairs of individuals. The hurdle model allows higher probabilities of no contact in a way similar to the use of a zero-inflated Poisson model. They used control limits based on a Bayesian predictive distributions on the contacts between each pairs of individuals to identify a subset of potentially anomalous pairs of individuals. If an observed count is sufficiently far into the tails of the predictive distribution, as measured by a *p*-value, a signal is given that there could be an anomaly. The predictive distribution for the current count was based on the prior distribution and all data up to, but not including, the current time. They then used standard network inference tools on a smaller sub-network based on the pairs of individuals identified as anomalous and their contacts to identify anomalous network behavior. Heard et al. (2010) used a *p*-value threshold of 0.05, which will lead to many pairs of individuals falsely identified as anomalous in large networks.

Heard et al. (2010) did not realize that a number of researchers have proposed using control charts with the control limits based on Bayesian predictive distributions for quality control applications. These include Menzefricke (2002, 2007, 2010a, 2010b, and 2013), Bayarri and Garcia-Donato (2005), Saghir (2015), and Raubenheimer and van der Merwe (2015). The primary way in which these methods differ is that only Heard et al. (2010) and Bayarri and Garcia-Donato (2005) updated the posterior distribution of the parameter of interest using all prior data without a distinction between the retrospective analysis of baseline Phase I data and the on-going real-time monitoring in Phase II. In this sense, their approaches and the dynamic reference approach of Azarnoush (2016) are closely related to the use of the self-starting methods of Quesenberry (1991, 1995). The other researchers used predictive distributions based on only the fixed set of Phase I data to determine the posterior distribution of the parameter or parameters of interest. Heard et al. (2011) pointed out, however, that for a longer term view, local models should be fit within shorter blocks of time, i.e., a moving window version of their method should be used.

## *2.3 Scan methods*

A number of researchers have proposed what are referred to as scan-based network monitoring schemes. In a frequently cited paper, Priebe et al. (2005) proposed a method for detecting increases in communication levels based on the sizes of the $k^{th}$ order neighborhoods of each individual, where $k = 0$, 1, and 2. The degree of an individual was referred to as the size of the $0^{th}$ order neighborhood. Standardized statistics were calculated over time for each of the three metrics for each individual using a moving window of a specified length to establish the baseline mean and standard deviation. A lower bound of one was used for the estimates of the standard deviation to avoid signals for relatively small changes in network behavior. A lower bound of one for the estimated standard deviation is also used in the Early Aberration Reporting System algorithm for monitoring count data used by the U.S. Centers for Disease Control and Prevention in their BioSense program. See Hutwagner et al. (2003), Tokars et al. (2009), and Szarka et al. (2011) for more information.

With the Priebe et al. (2005) method, the maximum of the three standardized network metrics at each time period is taken over the set of individuals in the network. The signal rule is based on these maxima. These maximum values are themselves standardized based on the estimated mean and standard deviation of previous maxima calculated over a moving window and a signal is given whenever a maximum is further than five standard deviations from its estimated mean. Their method was applied retrospectively to an Enron e-mail network, but it was stated that their method can be used for prospective network monitoring. We anticipate, however, that this method will be able to detect quickly only very large network changes because of the use of the maximum of the standardized metrics. The standardized metrics corresponding to an individual could become quite large, for example, without causing the maximum value to take an unusually large value.

We note that the Priebe et al. (2005) method is not a scan method in the sense of Kulldorff (2001) or Joner et al. (2008). In these examples of more traditional scan methods, the monitoring statistics are based on counts in moving temporal or spatiotemporal windows. The network monitoring scan methods are instead based on maximum values of standardized deviations of metrics over moving windows, where the maximum is taken over all of the nodes in the network.

In their Bayesian method, Heard et al. (2010) updated the estimates of the baseline parameter values after each time period, whereas Priebe et al. (2005) based the comparison baseline on a moving window of observations. Using a moving window approach allows for the network behavior to evolve slowly over time without necessarily having a signal that a process change has occurred. Heard et al. (2010), on the other hand, incorporated all data into the estimates of the level of the process to which a metric based on the current sample is compared. The moving window approach seems more reasonable to us. One must keep in mind, however, that data reflecting undetected network changes become incorporated into the baseline with a moving window approach. This makes it more difficult to detect an anomaly that is not detected as soon as it occurs. In addition, moving window approaches will not continue to signal a sustained anomaly.

In his scan-based approach, Sparks (2015, 2016) first ranked the individuals in an attempt to have the more associated individuals closer to each other in the ranking. Once the ranking was made, a spatio-temporal scan approach was taken to identify any anomalous sub-networks with increased communication levels. One advantage given for the approach is its computational efficiency compared to the infeasible approach of scanning over the activity of all subsets of individuals of given sizes. One concern with this approach is that communities within the network may not be captured by the ordering of the individuals. In addition, the network change to be detected may correspond to sub-networks different from those captured by the ordering of individuals.

Other scan-based approaches pointed out by Savage et al. (2014) include a dissertation by Neil (2011), the ideas in which were subsequently published in Neil et al. (2013). The application was on computer network monitoring, however, not social network monitoring. Other work involving scan statistics included Cheng and Dickinson (2013), Marchette (2012), Park et al. (2009), and McCullough and Carley (2011). The approach of Marchette (2012) is closely related to that of Priebe et al. (2011). McCullough and Carley (2011) claimed to use a scan approach similar to that of Priebe et al. (2005), but we find their scan method to be somewhat ambiguously defined.

Cheng and Dickinson (2013) combined the scan methods of Priebe et al. (2005) with an analysis of cross-correlations between the network metrics being monitored. The cross-correlations were calculated based on the data in the moving window, which does not include the most current observation. One concern regarding this approach is that the cross-correlations do not necessarily provide any information on anomalous activity that occurs with the current network graph. Also, their use of average correlations can mask important relationships between pairs of metric time series. Finally, it does not seem that they account for the fact that a correlation of a time series with itself will always be unity.

## 4. Time series models

Savage et al. (2014) mentioned Pincombe (2005) as providing a network monitoring method based on time series models. Time series models can be fitted to time series of any network metrics. Unusually large residuals indicate network changes. It is important to note that this type of approach has been widely used for process monitoring in public health surveillance and in industrial and quality-related applications. Woodall and Montgomery (2014) provided an overview of this area and cited several review papers on the use of time series models in process monitoring, including Psarakis and Papaleonida (2007). Unkel et al. (2011) reviewed the use of time series approaches in public health surveillance.

## 5. Other approaches

Many methods have been proposed in the network analysis literature for detecting changes in network structure or behavior over time with specific goals in mind. Examples include detecting fraudulent accounts, detecting unusual events affecting network behavior and detecting change in community structure. A complete review of these methods is not feasible, but we briefly discuss some of this work in this subsection.

Cazabet et al. (2010), for example, proposed a method for identifying changes in community structure over time where the identified communities could possibly overlap. As data are obtained, previously identified communities are updated and new communities can be identified.

As another example, Chae et al. (2012) proposed a method for detecting abnormal events quickly, such as a mass shooting or an earthquake, using social media data that incorporates spatiotemporal information. The approach involves a seasonal trend decomposition in conjunction with control chart methods based on a moving window of values to find unusual peaks and outliers within topic time series. In a related paper, Altshuler et al. (2013) developed a method for detecting an extraordinary event using the timing and traffic within a network assuming no knowledge of the content of the messages.

In addition, Egele et al. (2013) proposed a method for identifying compromised user accounts by building behavioral profiles for the users. Their method involves looking for groups of accounts that all experience similar changes within a short period of time. Their method was illustrated using Twitter and Facebook datasets. Takahashi et al. (2011), on the other hand, proposed a method for detecting emerging topics from social network streams based on the mentioning behavior of the users.

## 3. Some issues in social network monitoring

### *1. Phase I vs. Phase II*

In statistical process monitoring it is important to distinguish between Phase I and Phase II. Phase I includes methods for understanding process behavior based on a fixed baseline set of data. In-control parameter values for appropriate models are estimated in the retrospective Phase I and used to design methods for on-going prospective monitoring in Phase II. In Phase II, we make a decision about the stability of the process relative to the Phase I baseline as each sample is collected over time. Phase I issues and methods were discussed by Jones-Farmer et al. (2014).

Generally it would seem to be more difficult to obtain a baseline of stable network data, however, than it would be to obtain such data in a much more controlled industrial environment. Thus we see a greater need for the use of moving window approaches which would be inappropriate for industrial process monitoring because industrial processes are not allowed to wander or evolve.

Savage et al. (2014) referred to methods of network anomaly detection as being either "static" or "dynamic". For static network methods the time order of contacts is ignored with all data aggregated over time. We consider it useful to also distinguish between Phase I dynamic methods to be used on a set of historical data with time order preserved and Phase II dynamic monitoring performed on-line as each new matrix of counts is observed. Generally the methods used for the analysis of Phase I data differ from those used in Phase II. Quick detection of process changes is important in Phase II, for example, but irrelevant in the analysis of Phase I data. Thus EWMA and CUSUM methods are often used in Phase II, while change-point and outlier detection methods are commonly used in Phase I.

### *3.2 Use of computer simulation*

We agree with Savage et al. (2014) that methods need to be compared based on simulated networks. McCulloh and Carly (2011) also pointed out the usefulness of simulation studies. Anomalies can be modeled in the simulated datasets and methods can be compared on the basis of their ability to detect the anomalies. There is a substantive literature in the statistical modeling of networks that offers a diverse number of random graph models that may be helpful in this endeavor. For example, see the recent review of Goldenberg et al. (2010). There are advantages in using parametric statistical models for the networks so that multiple graphs can be generated to represent a baseline and so that anomalies can be simulated by changing the parameters corresponding, for example, to contacts between individuals within a sub-network. Ideally one should use realistic networks, but the use of simplified networks would likely provide valuable insights on the relative performance of competing methods. If a method is not effective in detecting changes in simple networks, it will be unlikely to be effective with more complex networks. Decisions are required on the number of individuals in the network, the grouping of individuals into sub-networks, the type of covariate information, if any, and the type of anomaly to be detected.

In their simulation Azarnoush et al. (2016) assumed a given logistic regression model for the probabilities of contacts between pairs of individuals. They assumed that covariate data was available on the individuals, i.e., the data were labelled. Miller et al. (2013) also used simulation to study the detection performance of their method. In his simulations Sparks (2015, 2016) assumed that the numbers of contacts between individuals were Poisson distributed.

## *3.3 Distributional assumptions*

To model a network parametrically requires some distributional assumptions. It is sometimes assumed that the number of communications between pairs of individuals is Poisson distributed. The Poisson means can vary depending on the sub-group membership of the individuals. See, for example, Sparks (2015, 2016). Heard et al. (2010) used a hurdle variant of the Poisson distribution to account for an increased probability of no communication between two individuals in a given time period. Savage et al. (2014) stated, however, that social network communication count distributions typically have heavier tails than those associated with the Poisson distribution. The use of Bayesian models can yield negative binomial distributions for the counts. The negative binomial distribution, frequently used in public health surveillance, can be used to model counts that are overdispersed relative to the Poisson distribution.

Poisson distributed numbers of contacts for individuals result from the random graph approach of Erdős and Rényi (1960) under the assumption that contact between any two specified individuals can be represented by a Bernoulli random variable with a constant probability. As pointed out by Miller at al. (2013), the degree distribution follows a power law distribution for many networks, in which case scale-free random graph models such as the preferential attachment model of Barabási and Albert (1999) can be used. Another option is the degree-corrected stochastic block model of Karrer and Newman (2011).

In their computer simulations Miller et al. (2013) and Azarnoush et al. (2016) assumed that there was covariate information on the individuals in the network. The probability of a link between any two individuals was modeled using log-linear modeling and logistic regression, respectively, in their approaches.

We do not support the use of the binomial model by Vigliotti and Hankin (2015). They proposed breaking each of time periods for which we are obtaining the matrices $C_t$ into disjoint increments, assuming that the probability of at least one connection between two individuals in each increment is a fixed value π. Thus, the sum of these Bernoulli random variables is a binomial random variable. The issues regarding how to divide the interval into increments and the estimation of π, however, were not addressed. In addition, if more than one contact is made between individuals in a single time increment, there would be a loss of information with their approach.

To simulate networks with parametric models, some assumptions about dependence structure are needed. As a start, it seems reasonable to assume independence of the $C_t$ matrices over time. If a method works poorly under this assumption, it would be unlikely to work well under a more general model.

### 3.4 *Performance metrics for monitoring schemes*

We require metrics in order to compare the performance of network monitoring methods in computer simulation studies. The standard performance metrics in quality control applications are based on the run length distribution, where the run length is the number of samples of observations until a signal is given that a process change has occurred. Typically the average run length (ARL) is used. One would like for the ARL to be suitably large when the process is stable and low when a process change occurs. McCullough and Carley (2011) defined an average detection length metric that is equivalent to the ARL.

The ARL metric is useful when a change in the process is sustained until it is detected. If a change to the network is temporary, however, then a more reasonable metric is the probability of detecting the process change while it is in effect. This is referred to as the probability of correct detection. A general discussion of this and other performance metrics was given by Frisén (1992) and Fraker et al. (2008).

In assessing performance in detecting a process change, it can be assumed that the process change happens at the time monitoring begins or that the change is delayed. Metrics under these two scenarios are referred to as being zero-state and steady state, respectively. Generally, steady-state performance metrics are preferred in statistical process monitoring because process changes are frequently delayed and because some methods have good zero-state performance, but poor steady-state performance. See, for example, Sego et al. (2008). We expect that the performance of the method of Miller et al. (2011, 2013) will not be as good for delayed network changes as it is for network changes that occur when monitoring begins.

In addition to quick detection of network anomalies, the individual or individuals involved in the anomaly may need to be accurately identified. This is analogous to being able to identify the correct geographical region of an outbreak in public health surveillance applications. Appropriate metrics include the percentages of misclassified individuals or the use of a metric such as the Dice similarity coefficient proposed by Dice (1945) and used by Megahed et al. (2012) in an image monitoring application. It may also be important to determine the time at which an anomaly first occurred. Amiri and Allahyari (2011) reviewed the statistical process monitoring literature on identifying the time of a process change after a signal that a change has occurred.

With large networks methods may tend to identify one or more individuals or sub-networks as being anomalous at each time period. In these cases the ARL metric is no longer useful. Metrics such as the false discovery rate would then be needed based on the ideas of Benjamini and Hochberg (1995) and Benjamini and Yekutieli (2001).

We note that the use of performance metrics is required in order to compare the performance of competing methods in simulation studies. Practitioners, however, should not expect to be able to design monitoring methods such that performance metrics will take specified values, e. g., having an in-control ARL of 100. As illustrated by Saleh et al. (2015), it is not possible to have enough baseline data to accomplish this objective even in the much simpler univariate case of monitoring the mean of a variable assumed to have a normal distribution.

## 4. Research opportunities and conclusions

We believe that the monitoring of social networks is an important application and research area with abundant opportunities available. We agree with McCullough and Carley (2011) that social network change detection represents an exciting new area of research.

The following are some research topics of interest:

1. We agree with Savage et al. (2014) that there is a need to evaluate and compare the performance of existing methods. As they point out, most authors simply illustrate their proposed methods based on case study datasets. One cannot reliably compare performance of methods based on case study results since one rarely knows whether or not any detection is a false alarm. In addition, a method tailor-made for a specific case study may perform poorly in other applications.

    Comparisons of existing methods would likely spark ideas for new methods. It is better if new methods are scalable to large networks.

2. We also agree with Savage et al. (2014) that research is needed to provide guidance on the selection of the most effective network metrics to monitor in order to satisfy the objectives of the monitoring.

3. Many of the approaches used are of the Shewhart-type in that the decision whether or not an anomaly is present is based on each set of graph information individually as it is obtained. See, for example, Azarnoush et al. (2016). McCulloh and Carley (2008a, b) advocated use of CUSUM and EWMA methods based on network metrics. We would expect that the CUSUM and EWMA methods would have better detection capability, but performance comparisons are needed.

4. Study is needed on the effect of aggregation over time on the monitoring of networks. This would be a generalization of the work of Schuh et al. (2013). We expect that detection of anomalies will become more difficult with increasing levels of aggregation, especially with Bernoulli data. In addition, study is needed on the effect of the loss of information in considering Bernoulli data instead of the numbers of contacts between individuals. We anticipate that reducing count data to Bernoulli data could result in a significant loss of information and a greatly reduced ability to detect network anomalies, particularly as graph data are aggregated over longer time intervals.

5. Is it more efficient to identify individuals with anomalous behavior and then analyze the resulting sub-network (as in Heard et al., 2010) or is it better to search over sub-networks directly by monitoring $k^{th}$ order neighborhood data corresponding to each individual (as in Priebe et al., 2005)? We anticipate that the latter approach will be more effective because the structure of the sub-network formed by individuals with anomalous behavior may not necessarily be anomalous.

6. We encourage further investigation of monitoring methods based on monitoring the eigenvalues of modularity matrices. It is important to clarify what types of network changes are not detectable with use of a specified number of eigenvalues.

7. The use of false discovery rate approaches seems appropriate for methods based on the simultaneous use of many charts, such as the method proposed by Heard et al. (2010). Woodall and Montgomery (2014) listed several papers on the use of the false discovery rate approach in process monitoring. In addition, see Gandy and Lau (2013). Some of the network monitoring methods, for example those by Heard et al. (2010) and Vigliotti and Hankin (2015), are already *p*-value based with a concern over the high number of false positives so use of a false discovery rate approach seems promising.

8. Additional methods are needed that incorporate covariate information about the network or the contacts. This could include labels that categorize individuals into groups, the length or size of the message constituting the contact, and the time of any contact. Savage et al. (2014) referred to the monitoring in this case as a search for dynamic labelled anomalies. Miller et al. (2013) and Azarnoush et al. (2016) seem to be the only ones thus far to have proposed methods for monitoring with attributed (or labelled) data.

9. Most often the graph count data are not smoothed over time. Moving window methods are used instead. Sparks (2015, 2016), however, smoothed the count data using exponential smoothing to build in temporal memory. It is not clear which approach is better.

10. With moving window approaches, what should the length of the window be in a given application? Azarnoush et al. (2016) used moving windows of sizes 4 and 10 whereas Priebe et al. (2005) used a window lengths of size 20. Also it seems that it may be possible to improve performance by lagging the window by not including a specified number of the most recent graphs.

11. There will likely be seasonal effects in network data, e.g., day of the week effects or holiday effects. Seasonal effects could be identified using Phase I data. Sometimes the effect of this variation can be removed by aggregating over the data over time, e.g., aggregation of daily data by weeks. Seasonal effects are common in public health monitoring applications, so some public health surveillance methods could likely be adapted for use with network data.

12. Methods must be adapted for evolving networks to account for new individuals entering the network and for individuals leaving the network. These events can trigger signals of network change that are not likely of interest.

13. As a quality monitoring research topic, a comparison is needed between Bayesian control charts based on predictive distributions and the self-starting control chart approaches.

## Acknowledgements

The work of W. H. Woodall was partially supported by National Science Foundation Grant CMMI-1436365.

Bayarri, M. J. and Garcia-Donato, G. G. (2005) A Bayesian sequential look at *u*-control charts. *Technometrics*, **47**(2), 142-151.

Benjamini, T. and Hochberg, T. (1995). Controlling the false discovery rate: A practical and powerful approach to multiple testing. *Journal of the Royal Statistical Society – Series B*, **57**, 289-300.

Benjamini, Y. and Yekutieli, D. (2001) The control of the false discovery rate in multiple testing under dependency. *Annals of Statistics*, **29**, 1165-1188.

Box, G. and Ramirez, J. (1991) Cumulative score charts. *Quality and Reliability Engineering International*, **8**, 17-27.

Cazabet, R., Amblard, F., and Hanachi, C. (2010) Detection of overlapping communities in dynamical social networks. IEEE International Conference on Social Computing/IEEE International Conference on Privacy, Security, Risk and Trust, 309-314.

Chae, J., Thom, D., Bosch, H., Jang, Y., Maciejewski, R., Ebert, D. S. and Ertl, T. (2012) Spatiotemporal social media analytics for abnormal event detection and examination using seasonal-trend decomposition. IEEE Conference on Visual Analytics Science and Technology (VAST), 143-152.

Cheng, A. and Dickinson, P. (2013) Using scan-statistical correlations for network change analysis. PAKDD 2013 Workshops, J. Li et al. (Eds.), LNAI 7867, 1-13.

Dice, L. R. (1945) Measures of the amount of ecologic association between species. *Ecology*, **26**, 297–302.

Egele, M., Stringhini, G., Kruegel, C., and Vigna, G. (2013) COMPA: Detecting compromised accounts on social networks. In Symposium on Network and Distributed System Security (NDSS).

Erdős, P. and Rényi, A. (1960) On the evolution of random graphs. *Publications of the Mathematical Institute of the Hungarian Academy of Sciences*, **5**, 17–61.

Fraker, S. E., Woodall, W. H., and Mousavi, S. (2008) Performance metrics for surveillance schemes. *Quality Engineering*, **20**, 451-464.
20

# APPENDIX: SHORT TUTORIAL ON NETWORK TERMINOLOGY

In this section, we present some common network terminology relevant to social network monitoring and analysis. In the context of social monitoring, networks provide a natural means to model the communication patterns among a group of individuals. A *network*, or *graph*, $G = (V, E)$ is a mathematical object with two major components: a *vertex set V* where each *vertex,* or *node,* represents an individual, and an *edge set E* that is a subset of $V \times V$, that contains all pairs of vertices $(i, j)$ such that there is an *edge* between nodes $i$ and $j$. Figure A.1 illustrates a small network with 12 nodes and 16 edges. In social networks an edge exists between two individuals forming a link provided there is at least one contact between them or some other criterion on the communication level is met. Information about the edges is contained in the adjacency matrix.

The nodes or the edges of a graph may be *labeled* so that a label specifies some quantitative or qualitative attribute for a node or edge, e.g., the number of contacts between the individuals. The nodes themselves could be labeled with names or e-mail addresses. Edges of an *unlabeled* graph for a social network contain no information other than the presence of a link. If the graph is *directed,* then relationships are asymmetric and an edge $(i, j)$ represents a directed communication from individual $i$ to individual $j$. If the graph is *undirected,* then which individual initiated the communication is unspecified. A graph is said to be simple if it does not contain multiple edges between vertices or any edge that starts and ends at the same node. The *order* of the graph *G* is the number of vertices *n*, and the *size* of *G* is the number of edges that it contains, denoted by |*E*|. When the communication between two individuals is discrete-valued, one also specifies a collection of *edge weights* {*w*(*i*, *j*): $1 \leq i, j \leq n$} so that *w*(*i*, *j*) represents the number of communications between individuals *i* and *j*.

The *degree* of a node *i* is the number of edges incident to *i*, i.e. the number of nodes with communications involving node *i*. The degree of a node *i* is also referred to as the *degree centrality* of that node.

Simple directed graphs with *n* nodes can contain a total of *n*(*n*-1) possible edges; whereas, simple undirected graphs with *n* nodes can contain only possible edges. The *edge density* of a graph is the ratio of the size of the graph to the total number of possible edges

A *sub-network* or *sub-graph* $G^s = (V^s, E^s)$ of *G* is a graph whose vertex set $V^s$ is a subset of *V* and whose edge set $E^s$ contains all edges shared among the vertices in $V^s$. In the context of social network monitoring, one often seeks a sub-graph of an observed graph at some time *t* such that the vertices in the sub-graph have a significantly increased or decreased rate of communication at time *t*. In Section 2 we reviewed methods for detecting such network changes.

A special case of a sub-graph is an *ego-net,* which consists of a particular node (called an *ego*), and the nodes (called *alters*) that are connected to the ego, as well as any edges among the alters. For example in Figure 3, the nodes {3, 4, 6} and the three edges connecting them forms an ego-net for node 3. This ego-net is also a *clique*, a sub-network in which there is at least one contact between all pairs of individuals in the sub-network.

In many cases, we are interested in the paths between two nodes in a graph. The *shortest path* between two nodes *i* and *j* is the collection of vertices and edges such that there is a *path* from *i* to *j* along the collection and the collection contains the fewest number of edges. The *closeness centrality* of a node *i* quantifies how close *i* is to the remainder of the graph using shortest paths. In particular, it is the inverse of the sum of the shortest distances from *i* to all other nodes in the graph. In a directed graph, the *minimum directed path length* between nodes *i* and *j* is the number of directed edges in the shortest path between *i* and *j*. For example, the minimum directed path length between nodes 2 and 9 in Figure 3 is two as the shortest path between the two nodes is 2 - 8 - 9. *Betweenness centrality* for a particular node is the average of the proportions of shortest paths between pairs on nodes that include the node of interest. *Eigenvector centrality* is another measure of the influence of a node in a network. It assigns relative scores to all nodes in the network based on the concept that connections to high-scoring nodes contribute more to the score of the node in question than equal connections to low-scoring nodes. Google's PageRank is a variant of the eigenvector centrality measure (Page et al., 1999). The various network metrics can be calculated for each individual or averaged over sub-networks or averaged over the entire network. These metrics can be monitored over time as discussed in Section A.1.

In undirected graphs, the *transitivity* of a graph is the ratio of the number of closed triplets of vertices to the total number of triplets. Note that a *triangle* contains three closed triplets. In Figure A.1, nodes {3, 4, 6} form a closed triplet; whereas, nodes {10, 0, 1} form an open triplet. The *neighborhood* of a node $i$ is the collection of vertices that share an edge with $i$. More generally, the *kth order neighborhood* of a node $i$ is the collection of vertices within $k$ edges of $i$. For example, in Figure A.1 the 2$^{nd}$ order neighborhood of node 0 is the collection {1, 7, 8, 9, 10}.

Figure 2 illustrates an example of a labeled, undirected network whose structure changes over time. This figure shows three snapshots of Enron email networks in (a) 2000, (b) 2001, and (c) 2002. Circles represent those in managerial levels, squares represent other employees and traders, and the rectangle shows the in-house lawyer. Different colors also represent different roles in the company. As can be seen from Figure 2(b), the email communication network became denser in 2001 when the Enron scandal occurred. Also the level of email communications between the in-house attorney and managers significantly increased in 2001.

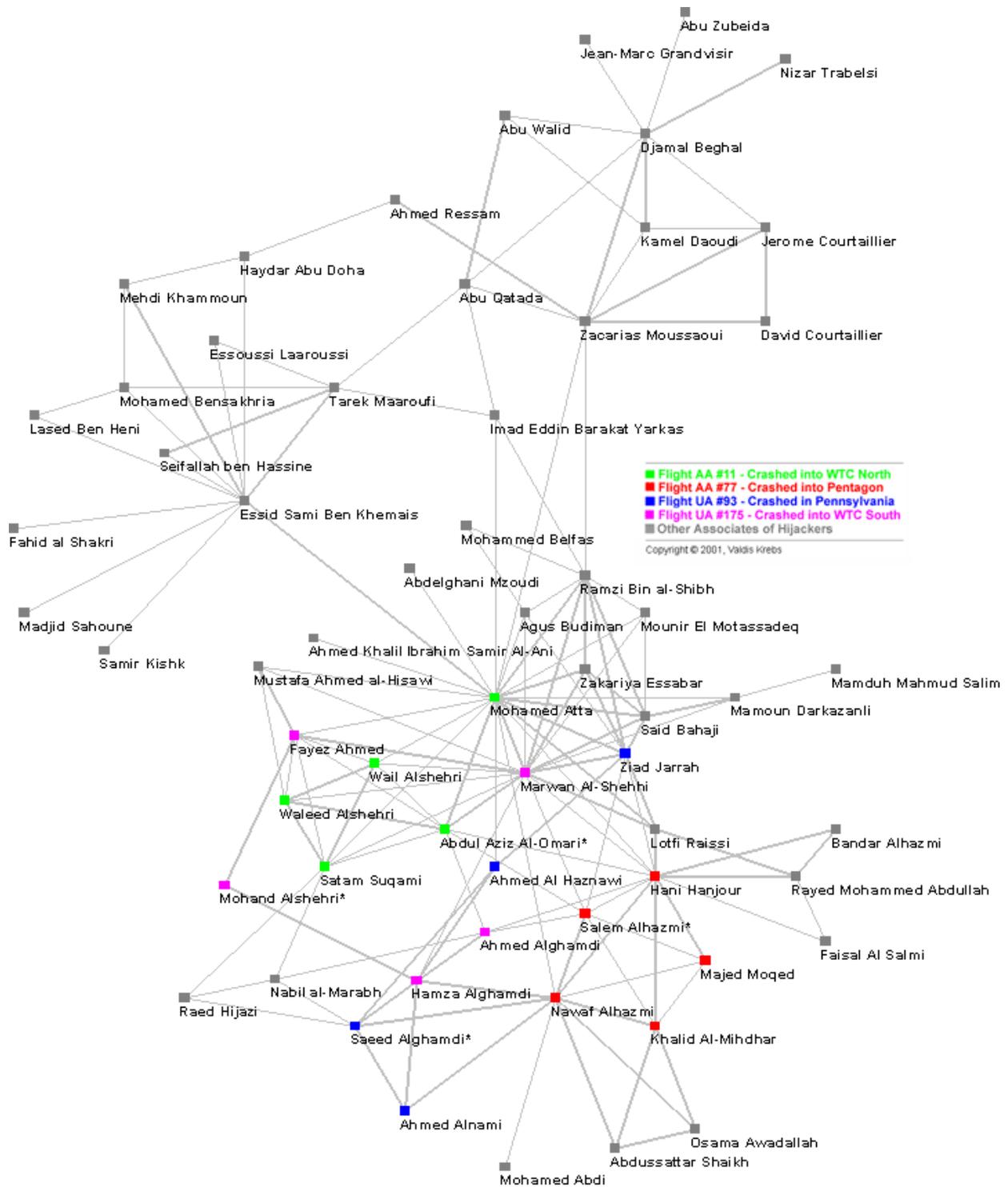

Figure 1. Illustration of terrorist network. From Krebs (2002).

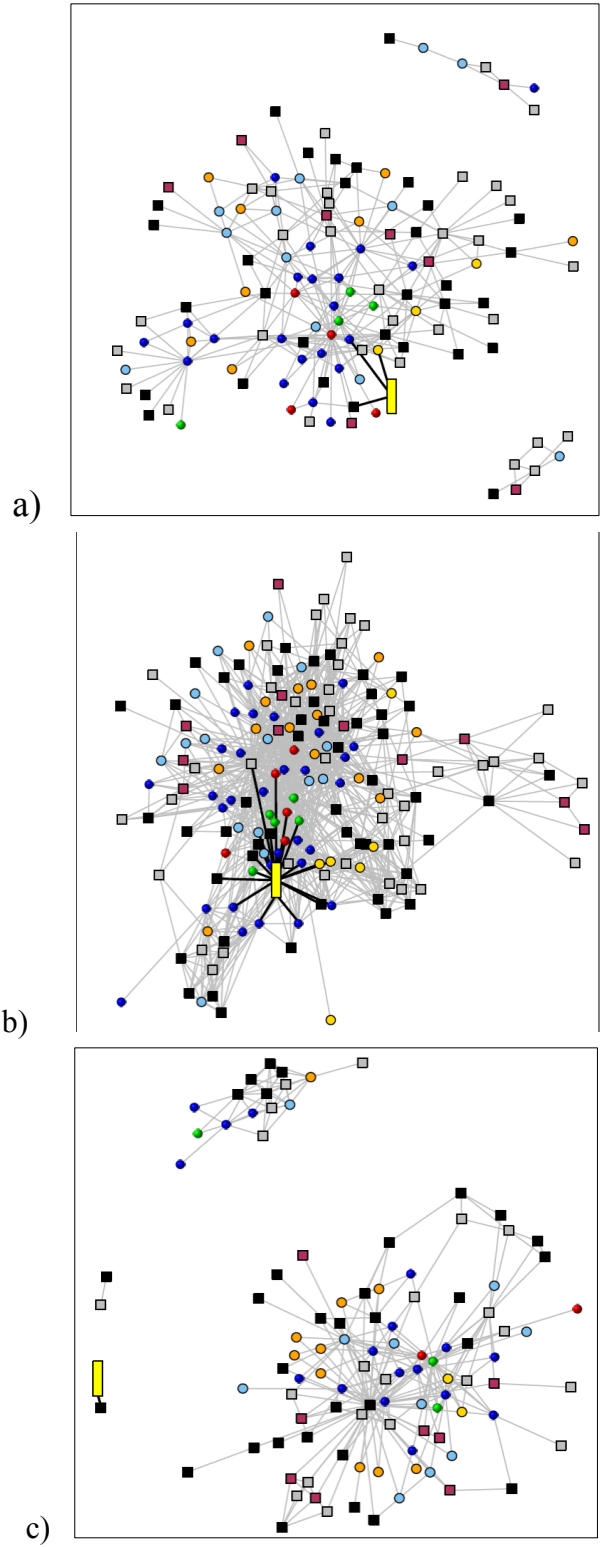

Figure 2. Illustration of three snapshots of Enron email networks in 2000 (plot a), 2001 (plot b), and 2002 (plot c).

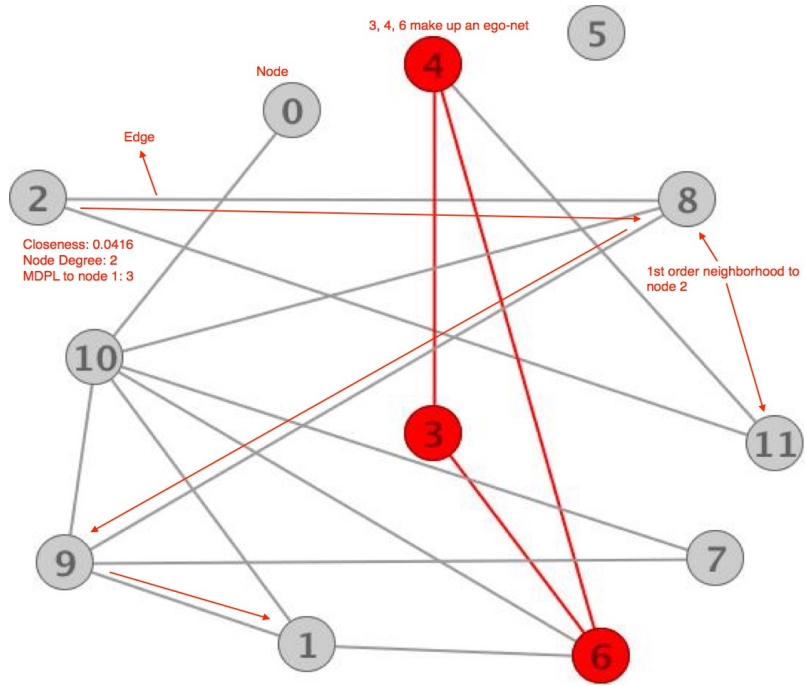

Figure A.1. Illustration of a small network with 12 nodes and 16 edges. Created using GraphTea software.

William H. Woodall is a Professor in the Department of Statistics at Virginia Tech. He is a former editor of the *Journal of Quality Technology* (2001–2003) and associate editor of *Technometrics* (1987–1995). He is the recipient of the Box Medal (2012), Shewhart Medal (2002), Jack Youden Prize (1995, 2003), Brumbaugh Award (2000, 2006), Søren Bisgaard Award (2012), Ellis Ott Foundation Award (1987), Lloyd S. Nelson Award (2014), and best paper award for IIE Transactions on Quality and Reliability Engineering (1997). He is a Fellow of the American Statistical Association, a Fellow of the American Society for Quality, and an elected member of the International Statistical Institute.

Meng J. Zhao is currently a Ph.D. candidate in the Department of Statistics at Virginia Tech. He is expected to receive his doctorate in statistics in 2017. He received a M.S. in statistics in the Department of Statistics at Virginia Tech and a M.S. in biotechnology at Pennsylvania State University. His Ph.D. research focuses in areas of statistical process monitoring and social network research. Prior to pursing a Ph.D. in statistics, he worked as a biochemist in the pharmaceutical industry for four years. He is a student member of American Statistical Association.

Kamran Paynabar is an Assistant Professor in the H. Milton Stewart School of Industrial & Systems Engineering at the Georgia Institute of Technology. He received B.Sc. and M.Sc. in Industrial Engineering from Iran University of Science and Technology and Azad University in 2002 and 2004, respectively, and Ph.D. in Industrial and Operations Engineering from The University of Michigan in 2012. He holds an M.A. in Statistics from The University of Michigan. His research interests include high-dimensional data analysis for systems monitoring, diagnostics and prognostics, and statistical and machine learning for complex-structured streaming data including multi-stream signals, images, videos, point clouds and network data He is interested in applications ranging from manufacturing, including automotive and aerospace, to healthcare. Dr. Paynabar is a member of Institute of Industrial Engineers (IIE) and The Institute for Operations Research and the Management Sciences (INFORMS).

Ross S. Sparks is a Team Leader of Real-time Modelling and Monitoring at Data61, CSIRO, Australia. He holds a B.Sc. with majors in Mathematics and Mathematical Statistics from University of Natal (1972), a B.Sc. (Hons, 1978) and M.Sc. (1980) from the University of South Africa and Ph.D. (1984) in Statistics from University of Natal. His research interests include statistical quality control and improvement, all aspects of process monitoring, prospective public health surveillance, applied multivariate analysis and applied statistics. He has lectured at University of Natal (1980-1983), University of Cape Town (1983-1988) and University of Wollongong (1988-1991) before joining CSIRO, Australia in 1991. Since 1991 he has worked at CSIRO on applied research projects for industry.

James D. Wilson is an Assistant Professor of Statistics in the Department of Mathematics and Statistics at the University of San Francisco. He holds an M.S. in Mathematical Sciences from Clemson University (2010), and a Ph.D. in Statistics and Operations Research from the University of North Carolina at Chapel Hill (2015). His research brings together techniques from machine learning, statistical inference, and random graph theory to model, analyze, and explore relational (network) data structures. His interdisciplinary work has led to collaborations with researchers from a variety of fields, including genetics, infectious disease, political science, and managerial science.